\documentclass[12pt,thmsa]{article}
\usepackage{sw20aip}

\input{tcilatex}
\begin{document}

\title{Asymptotic probability density functions in turbulence.}
\author{F. O. Minotti and E. Speranza \\
%EndAName
Instituto de F\'\i sica del Plasma, INFIP-CONICET\\
Departamento de F\'\i sica, Universidad de Buenos Aires,\\
1428 Buenos Aires, Argentina.}
\date{\today }
\maketitle

\begin{abstract}
A formalism is presented to obtain closed evolution equations for asymptotic
probability distribution functions of turbulence magnitudes. The formalism
is derived for a generic evolution equation, so that the final result can be
easily applied to rather general problems. Although the approximation
involved cannot be ascertained a priori, we show that application of the
formalism to well known problems gives the correct results.
\end{abstract}

\section{Introduction}

Probability distribution functions (pdf's) are powerful tools for the
experimental and theoretical study of turbulence. Experimental pdf's can be
obtained rather directly from the measurements of different magnitudes, and
are at the same time a very convenient tool for the representation of such
results. From the theoretical point of view, the equations for the evolution
of pdf's can be readily obtained from the basic dynamical equations,
although generally not in closed form. Closure requires the determination of
conditional averages in terms of the pdf itself, which sometimes can be done
based on additional information given by experiments\cite{chicago,poping},
and numerical simulations\cite{kraichen,chingcia}. Closures of this kind
have allowed to derive very important results directly from the original
dynamical equations, such as the limiting and time dependent pdf of passive%
\cite{sinai,val} and non-passive\cite{yakhot} scalars. We present here a
procedure for closure of the pdf equation, at least of its asymptotic form,
that can be applied in a systematic way to general systems. It is important
to mention that one can only show that the model is applicable if a
reasonable conjecture is valid. We show that the model gives correct results
in various systems whose behavior is known. The advantage is that it is very
easy to apply and so it is valuable to attack new problems for which no
useful information exists.

\section{Formalism}

The basic idea is most easily introduced considering a generic scalar field $%
\phi (x,t)$, where $t$ is the time, and $x$ denotes the spatial coordinates.
The field $\phi $ satisfies a generic evolution equation of the form 
\begin{equation}
\frac{\partial \phi }{\partial t}=N\left[ \phi ,x\right] +f\left( x,t\right)
,  \label{ns}
\end{equation}
in which $N\left[ \phi \right] $ is a spatial functional of $\phi \left(
x,t\right) $ evaluated at the actual time $t$, and $f$ is a stochastic
forcing introduced to allow eventual consideration of statistically
stationary pdf's. $N\left[ \phi \right] $ contains in general linear and
non-linear terms, and we restrict the non-linear terms to entire powers of $%
\phi $, not necessarily local. The induced probability distribution of a
given functional of $\phi $, $G\left[ \phi \right] $, can be conveniently
represented as 
\begin{equation}
P\left( \xi \right) =\left\langle \delta \left( G\left[ \phi \right] -\xi
\right) \right\rangle ,  \label{pdfgen}
\end{equation}
in which $\left\langle ...\right\rangle $ means average over ensembles of
realizations, and $\delta $ is Dirac's delta function. The evolution
equation of $P\left( \xi \right) $ is then determined from Eq. (\ref{ns}) as 
\begin{equation}
\frac{\partial P}{\partial t}=-\frac{\partial }{\partial \xi }\left[
\left\langle \left. \frac{\delta G}{\delta \phi \left( x_{1}\right) }*\left[
N\left[ \phi ,x_{1}\right] +f\left( x_{1},t\right) \right] \right| \xi
\right\rangle \,P\right] ,  \label{dpdtgen}
\end{equation}
where $\left\langle \left. ...\right| \xi \right\rangle $ represents average
conditioned on $G\left[ \phi \right] =\xi $, and the asteric indicates
integration over repeated variables. We will consider only functionals $%
G\left[ \phi \right] $ that are linear in $\phi $, so that $\left( \delta
G/\delta \phi \right) *$ will be an operator independent of $\phi $ which we
denote as 
\[
\frac{\delta G}{\delta \phi \left( x_{1}\right) }\equiv h\left( x_{1}\right)
. 
\]
Besides, as $h\left( x_{1}\right) $ is independent of $\phi $, only $%
\left\langle \left. N\left[ \phi \right] +f\right| \xi \right\rangle $ needs
to be evaluated.

To evaluate $\left\langle \left. N\left[ \phi \right] \right| \xi
\right\rangle $ let us consider the simplest linear term, 
\[
\Gamma _{1}(y,\xi )\equiv \left\langle \phi (y)|h\left( x_{1}\right) *\phi
(x_{1})=\xi \right\rangle 
\]
where $y$ is an arbitrary spatial position and, from now on, no explicit
indication of the time is made. The exact, formal expression of $\Gamma
_{1}(y,\xi )$ can be written as 
\begin{equation}
\Gamma _{1}(y,\xi )=\frac{\int \mathcal{F}[\phi ]\phi (y)\delta \left(
h\left( x_{1}\right) *\phi (x_{1})-\xi \right) \mathcal{D}\phi }{\int 
\mathcal{F}[\phi ]\delta \left( h\left( x_{1}\right) *\phi (x_{1})-\xi
\right) \mathcal{D}\phi }.  \label{general}
\end{equation}
Here Dirac's delta function $\delta $ is used to select only those fields
that satisfy the condition $h\left( x_{1}\right) *\phi (x_{1})=\xi $, $%
\mathcal{F}[\phi ]$ is the probability density functional (pdF) of the field 
$\phi $, and the integrations are meant to be functional integrations over
fields $\phi $, with an appropriate measure $\mathcal{D}\phi $. In
principle, given the system (\ref{ns}) a Martin-Siggia-Rose lagrangian can
be determined in terms of which to express $\mathcal{F}[\phi ]$. Once this
is done, solutions of (\ref{general}) can be obtained in terms of
perturbative expansions, or non-perturbative approaches such as rapid
descent type of integrations around significant field configurations. We
follow here the perturbative approach, setting up an infinite series that
represents (\ref{general}) in a formally exact manner, and then dividing
this series into two infinite other series, such that one of them is term by
term much larger than the other in the limit of large $\xi $. We keep only
these larger terms and sum the resulting infinite series.

To proceed we write both integrands appearing in (\ref{general}) as series
involving Gaussian functionals. For this we write 
\begin{equation}
\mathcal{F}[\phi ]=\mathcal{F}_{0}[\phi ]\;\mathcal{G}[\phi ],  \label{fdec}
\end{equation}
where 
\begin{equation}
\mathcal{F}_{0}[\phi ]=Z_{0}^{-1}\exp [-\kappa (x_{1})*\phi (x_{1})-\frac{1}{%
2}\sigma (x_{1},x_{2})*\phi (x_{1})\phi (x_{2})],  \label{f0gauss}
\end{equation}
is a Gaussian functional, with $\kappa (x)$ and $\sigma (x,x^{\prime })$
functions to be determined, and $Z_{0}$ a normalization constant ensuring
that $\int \mathcal{F}_{0}[\phi ]\;\mathcal{D}\phi =1$. $\mathcal{G}[\phi ]$
is in general a non-Gaussian functional factor defined by the very
expression (\ref{fdec}), satisfying also the normalization condition $\int 
\mathcal{F}[\phi ]\;\mathcal{D}\phi =1$. We further expand $\mathcal{G}[\phi
]$ as a functional Taylor series 
\begin{equation}
\mathcal{G}[\phi ]=\sum\limits_{n=0}^{\infty }\frac{1}{n!}%
K(x_{1},...,x_{n})*\phi (x_{1})...\phi (x_{n}),  \label{gfunct}
\end{equation}
where the $n=0$ term is to be understood as constant.

We turn now to the evaluation of (\ref{general}). For this, note that the
denominator of this expression is the pdf (\ref{pdfgen}), $P(\xi )$, which,
using the Fourier representation of Dirac's delta function, can be written
as 
\begin{eqnarray}
P(\xi ) &=&\int \mathcal{F}[\phi ]\delta \left( h\left( x_{1}\right) *\phi
(x_{1})-\xi \right) \mathcal{D}\phi =  \nonumber \\
&&\ \ (2\pi )^{-1}\int dk\exp (-ik\xi )  \nonumber \\
&&\ \ \int \mathcal{F}_{0}[\phi ]\mathcal{G}[\phi ]\exp \left[ ikh\left(
x_{1}\right) *\phi (x_{1})\right] \mathcal{D}\phi .  \label{larga}
\end{eqnarray}
It is now useful to introduce an auxiliary functional given by 
\begin{eqnarray}
\mathcal{A}[J] &=&(2\pi )^{-1}\int dk\exp (-ik\xi )  \nonumber \\
&&\ \ \int \mathcal{F}_{0}[\phi ]\exp \left[ J\left( x_{1}\right) *\phi
(x_{1})+ikh\left( x_{1}\right) *\phi (x_{1})\right] \mathcal{D}\phi \mathbf{.%
}  \label{bfunct}
\end{eqnarray}
$\mathcal{A}$ plays the role of a potential from which (\ref{larga}) and the
numerator of (\ref{general}) can be obtained by differentiation as it is
immediately seen from (\ref{bfunct}) that, for instance, 
\begin{equation}
\frac{\delta \mathcal{A}}{\delta J(x_{k})}=\phi (x_{k})\mathcal{A}[J],
\label{dadjk}
\end{equation}
and so, successive derivatives allow to produce the factors $\phi $
appearing in the series of $\mathcal{G}[\phi ]$ (see (\ref{gfunct})). In
this way, we write the denominator and numerator of (\ref{general})
respectively as 
\begin{equation}
P(\xi )=\sum\limits_{n}\frac{1}{n!}K(x_{1},...,x_{n})*\left. \mathcal{A}%
^{(n)}\right| _{J=0},  \label{pb}
\end{equation}
and 
\begin{equation}
\Gamma _{1}(y,\xi )P(\xi )=\sum\limits_{n}\frac{1}{n!}K(x_{1},...,x_{n})*%
\left. \frac{\delta \mathcal{A}^{(n)}}{\delta J(y)}\right| _{J=0},
\label{gpb}
\end{equation}
where 
\begin{equation}
\mathcal{A}^{(n)}\equiv \frac{\delta ^{n}\mathcal{A}}{\delta
J(x_{1})...\delta J(x_{n})}.  \label{an}
\end{equation}
Evaluating the Gaussian integrals in (\ref{larga}) and (\ref{bfunct}) we
obtain 
\begin{equation}
\mathcal{A}[J]=N\exp \left( -\Delta ^{-1}\xi ^{2}/2\right) \exp W[J-\kappa ],
\label{apoten}
\end{equation}
where $N$ is a factor independent of $J$ and $\xi $, $\Delta =\Sigma
(x_{1},x_{2})*h(x_{1})h(x_{2})$, and 
\begin{equation}
W[X]=1/2\left[ \Sigma (x_{1},x_{2})*X(x_{1})X(x_{2})-\Delta
^{-1}b^{2}+2\Delta ^{-1}b\xi \right] ,  \label{wj}
\end{equation}
with $b[X]=\Sigma (x_{1},x_{2})*h(x_{1})X(x_{2})$ a linear functional of $X$%
, and $\Sigma $ is the inverse of $\sigma $ in the sense that 
\begin{equation}
\Sigma (x_{1},x_{2})*\sigma (x_{2},x_{3})=\sigma (x_{1},x_{2})*\Sigma
(x_{2},x_{3})=\delta (x_{1}-x_{3}).  \label{identi}
\end{equation}
In this way, (\ref{general}) can be calculated by explicit differentiation
through (\ref{pb}) and (\ref{gpb}), the series so obtained being formally
equivalent to the original integrals.

So far we have only set up a generic series expansion in analogy with usual
methods, and to obtain useful results this series needs to be evaluated at
least approximately. We assume no intrinsic small parameter, but rather
consider the condition of large values of $|\xi |$. Differentiating (\ref
{apoten}) one can write 
\begin{equation}
\frac{\delta \mathcal{A}}{\delta J(y)}=\frac{\delta W}{\delta J(y)}\mathcal{A%
},  \label{da2}
\end{equation}
We will not write at this point the explicit expression of $\delta W/\delta
J(y)$ but just point out its pertinent properties:

i) It is a linear functional of $J$ and a linear function of $\xi $.

ii) Every derivative of $\delta W/\delta J(y)$ with respect to $J$ deletes a
factor $\xi $. In contrast, every derivative of $\mathcal{A}$ brings up a
factor $\xi $.

With these considerations we now examine a generic derivative in the sum
over $n$ in (\ref{gpb}) and write in a notation symbolizing the $n$-order
derivative of the product in the r.h.s. of (\ref{da2}) 
\begin{equation}
\frac{\delta \mathcal{A}^{(n)}}{\delta J(y)}=\frac{\delta ^{n}\left( \delta 
\mathcal{A}/\delta J(y)\right) }{\delta J(x_{1})...\delta J(x_{n})}%
=\sum\limits_{p=0}^{n}\binom{n}{p} \mathcal{A}^{(n-p)}\frac{\delta ^{p}}{%
\delta J^{p}}\left[ \frac{\delta W}{\delta J(y)}\right] .  \label{snm}
\end{equation}
By property i) the sum over $p$ runs only from 0 to 1, so that we can write
the expansion (\ref{gpb}) as 
\begin{eqnarray*}
\Gamma _{1}(y,\xi )P(\xi ) &=&\left. \frac{\delta W}{\delta J(y)}\right|
_{J=0}\sum\limits_{n}\frac{1}{n!}K(x_{1},...,x_{n})*\left. \mathcal{A}%
^{(n)}\right| _{J=0} \\
&&+\sum\limits_{n}\left\{ \frac{1}{\left( n-1\right) !}K(x_{1},...,x_{n})*%
\right. \\
&&\left. \left[ \mathcal{A}^{(n-1)}\frac{\delta ^{2}W}{\delta J(x_{n})\delta
J(y)}\right] _{J=0}.\right\}
\end{eqnarray*}
By property ii) each ``$n$'' term in the first series is two orders in $\xi $
higher than the corresponding term in the second series and, besides,
according to expression (\ref{pb}) the first series is equal to $P(\xi )$.
In this way, the sum of the second series must be equal to $\left( \Gamma
_{1}(y,\xi )-\left. \delta W/\delta J(y)\right| _{J=0}\right) P(\xi )$,
which, by construction, is a finite magnitude. We have thus two convergent
series, one of which is term by term much larger than the other in the limit 
$|\xi |\rightarrow \infty $. Although this is not sufficient to ensure that
the sum of the first series dominates over the sum of the second series, we
obtain a plausible, simple model if we assume it and write in the large $%
|\xi |$ limit 
\begin{equation}
\Gamma _{1}(y,\xi )\rightarrow \left. \delta W/\delta J(y)\right|
_{J=0}=\Delta ^{-1}\xi \Sigma (y,x_{1})*h(x_{1}),  \label{gpfin}
\end{equation}
where $\Delta =\Sigma (x_{1},x_{2})*h(x_{1})h(x_{2})$. The function $\kappa $
does not appear in (\ref{gpfin}) because it contributes only to lower order
terms in $\left. \delta W/\delta J(y)\right| _{J=0}$.

The same argument applies to a generic non-linear term as 
\begin{equation}
\Gamma _{m}(y_{1},...,y_{m},\xi )\equiv \left\langle \phi (y_{1})...\phi
(y_{m})|h\left( x_{1}\right) *\phi (x_{1})=\xi \right\rangle ,  \label{gamam}
\end{equation}
which can be obtained from (compare with Eq. (\ref{gpb})) 
\[
\Gamma _{m}P(\xi )=\sum\limits_{n}\frac{1}{n!}K(x_{1},...,x_{n})*\left. 
\frac{\delta ^{m}\mathcal{A}^{(n)}}{\delta J(y_{1})...\delta J(y_{m})}%
\right| _{J=0}. 
\]
Properties i) and ii) allow this series to be split into $m+1$ series, such
that $m$ of them are term by term much smaller (by factors $\xi ^{-2}$, ..., 
$\xi ^{-2m}$) than the ``dominant'' series 
\[
\left. \frac{\delta ^{m}W}{\delta J(y_{1})...\delta J(y_{m})}\right|
_{J=0}\sum\limits_{n}\frac{1}{n!}K(x_{1},...,x_{n})*\left. \mathcal{A}%
^{(n)}\right| _{J=0}. 
\]
Again, by comparison with expression (\ref{pb}), this series is convergent,
so that the remaining series is also convergent. We thus model $\Gamma _{m}$
for large values of $\left| \xi \right| $ as 
\begin{eqnarray}
\Gamma _{m}(y_{1},...,y_{m},\xi ) &\rightarrow &\left. \frac{\delta ^{m}W}{%
\delta J(y_{1})...\delta J(y_{m})}\right| _{J=0}  \nonumber \\
&=&\ \Delta ^{-m}\xi ^{m}\Sigma (y_{1},x_{1})*  \nonumber \\
&&h(x_{1})...\Sigma (y_{m},x_{m})*h(x_{m}).  \label{gamammodel}
\end{eqnarray}

An important consistency check is that contraction of (\ref{gamammodel})
with $h(y_{1})...h(y_{m})$ satisfies the exact identity 
\[
\Gamma _{m}(y_{1},...,y_{m},\xi )*h(y_{1})...h(y_{m})=\xi ^{m}, 
\]
for any $h(x_{1})$.

So far no conditions have been imposed on the functions $\kappa (x)$ and $%
\sigma (x,y)$. The only restriction is that both, $\mathcal{F}_{0}[\phi ]$
and $\mathcal{F}[\phi ]$ are pdF's of $\phi $, so that they are normalized
to unity for the same measure $\mathcal{D}\phi $ 
\[
\int \mathcal{F}_{0}[\phi ]\;\mathcal{D}\phi =\int \mathcal{F}[\phi ]\;%
\mathcal{D}\phi =1, 
\]
and that, of course, $\mathcal{F}[\phi ]$ is independent of $\kappa (x)$ and 
$\sigma (x,y)$. From these conditions and the expression (\ref{f0gauss}) of $%
\mathcal{F}_{0}[\phi ]$ we easily obtain 
\begin{mathletters}
\label{dlng}
\begin{eqnarray}
\frac{\delta \ln \mathcal{G}}{\delta \kappa (x)} &=&-\frac{\delta \ln 
\mathcal{F}_{0}}{\delta \kappa (x)}=\phi (x)-\left\langle \phi
(x)\right\rangle _{0}, \\
\frac{\delta \ln \mathcal{G}}{\delta \sigma (x,y)} &=&-\frac{\delta \ln 
\mathcal{F}_{0}}{\delta \sigma (x,y)}=\frac{1}{2}\left[ \phi (x)\phi
(y)-\left\langle \phi (x)\phi (y)\right\rangle _{0}\right] ,
\end{eqnarray}
where $\left\langle ...\right\rangle _{0}$ stands for the average using $%
\mathcal{F}_{0}$. To choose the functions $\kappa (x)$ and $\sigma (x,y)$
consider that we have modeled the large $\xi $ asymptotics of the $\Gamma
_{m}$ terms as the sum of a ``dominant'' series in the sense described
above. Since the sums involve infinite series, the term by term dominance is
not sufficient to ensure dominance of the sum, and that is why what we
obtain is an approximation whose validity cannot be ascertained in general.
However, we expect the approximation to be better the faster the series (\ref
{gfunct}) converge, so that fewer terms are required in the sums to
approximate the functions at a given value of $\xi $. With only the freedom
to choose $\kappa (x)$ and $\sigma (x,y)$ we can then expect better results
if these functions are chosen so as to keep $\mathcal{G}[\phi ]$ as small as
possible in some averaged sense. Taking into account that $\mathcal{G}$ is a
definite positive magnitude, we could minimize $\int \mathcal{FG}\;\mathcal{D%
}\phi $, or $\int \mathcal{FG}^{2}\;\mathcal{D}\phi $, or $\int \mathcal{F}%
\ln \mathcal{G}\;\mathcal{D}\phi $, etc.. Of these possible choices, the one
that weights preferentially the small values of $\mathcal{G}$ is $\int 
\mathcal{F}\ln \mathcal{G}\;\mathcal{D}\phi $, which, using relations (\ref
{dlng}), leads immediately to ($\left\langle ...\right\rangle $ represents
the average using $\mathcal{F}$; that is, the true average)

\end{mathletters}
\begin{equation}
\left\langle \phi (x)\right\rangle =\left\langle \phi (x)\right\rangle
_{0}=-\Sigma (x,x_{1})*\kappa (x_{1}),  \label{phi1}
\end{equation}
\begin{eqnarray}
\left\langle \phi (x)\phi (y)\right\rangle &=&\left\langle \phi (x)\phi
(y)\right\rangle _{0}=  \nonumber \\
&&\ \ \ \ \ \ \Sigma (x,y)+\kappa (x_{1})\kappa (y_{1})*\Sigma
(x_{1},x)\Sigma (y_{1},y).  \label{phi2}
\end{eqnarray}
These equations are easily inverted to determine the functions $\Sigma
=\sigma ^{-1}$ and $\kappa $ as 
\begin{equation}
\Sigma (x,y)=\left\langle \phi (x)\phi (y)\right\rangle -\left\langle \phi
(x)\right\rangle \left\langle \phi (y)\right\rangle .  \label{sigmafin}
\end{equation}
\begin{equation}
\kappa (x)=-\sigma (x,x_{1})*\left\langle \phi (x_{1})\right\rangle ,
\label{kapafin}
\end{equation}

We then summarize the model for the asymptotic averages as 
\begin{eqnarray}
\Gamma _{m}(y_{1},...,y_{m},\xi ) &\rightarrow &\xi ^{m}\ \left[ \Sigma
(x_{0},x_{0}^{\prime })*h(x_{0})h(x_{0}^{\prime })\right] ^{-m}  \nonumber \\
&&\Sigma (y_{1},x_{1})*h(x_{1})...\Sigma (y_{m},x_{m})*h(x_{m}),
\label{finalmodel}
\end{eqnarray}
with $\Sigma (x,y)$ given by Eq. (\ref{sigmafin}).

Finally, the conditional average of the stochastic Gaussian force of zero
mean can be obtained explicitly for the case of very short lived force
correlation of the form\cite{nov} 
\[
\left\langle f(x,t)f(x^{\prime },t^{\prime })\right\rangle =\delta
(t-t^{\prime })\,F(x-x^{\prime }), 
\]
and is given in our notation as 
\begin{equation}
\left\langle \left. f\left( x,t\right) \right| \xi \right\rangle =-\frac{1}{2%
}\frac{\partial p}{\partial \xi }h\left( x_{1}\right) *F\left(
x-x_{1}\right) .  \label{forcing}
\end{equation}

\section{Applications}

\subsection{Temperature diffusion term}

As a first simple check, we consider the average related to temperature
diffusion for which measurements exist\cite{ching} 
\[
s(\xi )\equiv \left\langle \nabla ^{2}\phi |\phi (x)=\xi \right\rangle . 
\]
It corresponds to $h(x_{1})=\delta \left( x_{1}-x\right) $, and can be
modeled according to Eq. (\ref{finalmodel}) as 
\begin{eqnarray}
s(\xi ) &=&\left. \nabla _{y}^{2}\Gamma _{1}(y,\xi )\right| _{y=x}\rightarrow
\nonumber \\
&&\xi \frac{\left\langle \phi (x)\nabla ^{2}\phi \right\rangle -\left\langle
\phi (x)\right\rangle \nabla ^{2}\left\langle \phi (x)\right\rangle }{%
\left\langle \phi ^{2}(x)\right\rangle -\left\langle \phi (x)\right\rangle
^{2}}.  \label{qe}
\end{eqnarray}
It is convenient to take $\phi $ as a non-dimensional field of zero mean and
unit variance, defined from the physical temperature field $T(x)$ as 
\begin{equation}
\phi (x)=\frac{T(x)-\left\langle T(x)\right\rangle }{\sigma _{T}},
\label{nd}
\end{equation}
where $\sigma _{T}$ is the variance of $T$. The measured non-dimensional
conditional average is defined as 
\begin{equation}
r(\xi )\equiv \left\langle (\left| \nabla \phi \right| ^{2}\right\rangle
^{-1}s(\xi ),  \label{re}
\end{equation}
which, from (\ref{qe}) and the fact that by definition the variance of $\phi 
$ is constant, reduces to 
\begin{equation}
r(\xi )\rightarrow -\xi .  \label{rehom}
\end{equation}
This simple result of the model has been experimentally seen to hold for a
wide range of values of $\xi $ \cite{poping}.

\subsection{Forced Burgers turbulence}

For a randomly forced Burgers flow, the velocity $u(x,t)$ satisfies\cite{bur}
\begin{equation}
\frac{\partial u}{\partial t}+u\frac{\partial u}{\partial x}=\nu \frac{%
\partial ^{2}u}{\partial x^{2}}+f(x,t),  \label{bs}
\end{equation}
where $\nu $ is the kinematic viscosity, $f(x,t)$ is a random force with
Gaussian distribution of zero mean and variance given by 
\begin{equation}
\left\langle f(x,t)f(x^{\prime },t^{\prime })\right\rangle =\delta
(t-t^{\prime })\,F(x-x^{\prime }),  \label{variance}
\end{equation}
with $F$ an even function of its argument which decays sufficiently fast for 
$\left| x-x^{\prime }\right| $ larger than a correlation length $L_{c}$.
Besides, $F(0)=2\varepsilon ^{\prime }$, where $\varepsilon ^{\prime }$ is
the rate of injection of energy density. Eq. (\ref{bs}) with forcing defined
by (\ref{variance}) has been extensively studied using different, powerful
theoretical methods\cite{pol,gur,bou,bal,wein,bol} and numerical simulations%
\cite{chekhot,yaklov}.

\subsubsection{Pdf of velocity difference}

The statistical magnitude to be considered first is the instantaneous
velocity difference across a separation $r$, $G[u]=u(x+r,t)-u(x,t)$. The
evolution equation for the corresponding pdf $P(\xi ,r,t)$ is, from Eqs. (%
\ref{pdfgen}) and (\ref{forcing}), 
\begin{equation}
\frac{\partial P}{\partial t}=\frac{\partial }{\partial \xi }\left(
\left\langle A|\xi ,r\right\rangle P\right) +\left[ F(0)-F(r)\right] \frac{%
\partial ^{2}P}{\partial \xi ^{2}},  \label{dpdt}
\end{equation}
where 
\begin{equation}
A=\left[ u(\eta ,t)\frac{\partial }{\partial \eta }u(\eta ,t)-\nu \frac{%
\partial ^{2}}{\partial \eta ^{2}}u(\eta ,t)\right] *h(\eta ),  \label{acond}
\end{equation}
$h(\eta )=\delta G/\delta u(\eta )=\delta (x+r-\eta )-\delta (x-\eta )$,
with $\left\langle ...|\xi ,r\right\rangle $, as above, a short notation for 
$\left\langle ...|u(x+r,t)-u(x,t)=\xi \right\rangle $. Finally, statistical
spatial homogeneity allows to simplify the expression of $A$ to write (\ref
{dpdt}) as 
\begin{eqnarray}
\frac{\partial P}{\partial t} &=&-2\frac{\partial }{\partial \xi }\left(
\left\langle u\frac{\partial u}{\partial x}|\xi ,r\right\rangle P\right) + 
\nonumber \\
&&\ \ \ \ \ 2\nu \frac{\partial }{\partial \xi }\left( \left\langle \frac{%
\partial ^{2}u}{\partial x^{2}}|\xi ,r\right\rangle P\right) +  \nonumber \\
&&\left[ F(0)-F(r)\right] \frac{\partial ^{2}P}{\partial \xi ^{2}}.
\label{dpfin}
\end{eqnarray}
Using the model (\ref{finalmodel}) we now evaluate the asymptotic
expressions of the conditional averages appearing in this equation, which
are conveniently expressed as 
\begin{eqnarray}
2\left\langle u\frac{\partial u}{\partial x}|\xi ,r\right\rangle &=&\left. 
\frac{\partial }{\partial R}\left\langle u^{2}(R)|\xi ,r\right\rangle
\right| _{R=0},  \label{r1} \\
\nu \left\langle \frac{\partial ^{2}u}{\partial x^{2}}|\xi ,r\right\rangle
&=&\nu \left. \frac{\partial ^{2}}{\partial R^{2}}\left\langle u(R)|\xi
,r\right\rangle \right| _{R=0}.  \label{r2}
\end{eqnarray}
Using expression (\ref{finalmodel}) with $m=1,2$ one obtains asymptotically 
\begin{equation}
\left\langle u(R)|\xi ,r\right\rangle \rightarrow \frac{\xi }{2S_{2}(r)}%
\left[ S_{2}(R)-S_{2}(r-R)\right] .  \label{s2fin}
\end{equation}
\begin{eqnarray}
\left\langle u^{2}(R)|\xi ,r\right\rangle &\rightarrow &\frac{\xi ^{2}}{%
4S_{2}^{2}(r)}\left[ S_{2}^{2}(r-R)+\right.  \nonumber \\
&&\ \ \ \ \ \left. S_{2}^{2}(R)-2\ S_{2}(R)S_{2}(r-R)\right] ,  \label{s1fin}
\end{eqnarray}
where it was used that in the spatially homogeneous case considered one can
write $\Sigma \left( x,x^{\prime }\right) =\left\langle u(x)u(x^{\prime
})\right\rangle =\left\langle u(x)u(x)\right\rangle -1/2S_{2}(x-x^{\prime })$%
, with $S_{2}(x-x^{\prime })=\left\langle \left( u(x)-u(x^{\prime })\right)
^{2}\right\rangle $ the second order structure function. To evaluate the $R$
derivatives appearing in (\ref{r1}) and (\ref{r2}) one needs to know the
behavior of $S_{2}$ for small separation. This can be done because at very
small scales, less than the dissipative scale set up by viscous effects, the
velocity field is a smooth function of $x$ and so one has\cite{stv,sir} $%
u(x+R)=u(x)+R\partial u/\partial x+O(R^{2})$, which leads to $%
S_{2}(R)=R^{2}\left\langle (\partial u/\partial x)^{2}\right\rangle
+O(R^{3}) $. With all this, (\ref{r1}) and (\ref{r2}) can be readily
evaluated as 
\begin{eqnarray}
2\left\langle u\frac{\partial u}{\partial x}|\xi ,r\right\rangle
&\rightarrow &-\frac{1}{2}\,\xi ^{2}\frac{\partial \ln S_{2}}{\partial r},
\label{duas} \\
\nu \left\langle \frac{\partial ^{2}u}{\partial x^{2}}|\xi ,r\right\rangle
&\rightarrow &\xi \frac{\varepsilon \Pi (r)}{S_{2}(r)}.  \label{d2uas}
\end{eqnarray}
where it was used that $\nu \left\langle (\partial _{x}u)^{2}\right\rangle
=\varepsilon $, with $\varepsilon $ the rate of dissipation of energy
density, and where $\Pi (r)=1-\nu \left( \partial ^{2}S_{2}/\partial
r^{2}\right) /(2\varepsilon )$. The asymptotic equation then reads 
\begin{eqnarray}
\frac{\partial P}{\partial t} &=&\frac{1\,}{2}\frac{\partial \ln S_{2}}{%
\partial r}\frac{\partial }{\partial \xi }\left( \xi ^{2}P\right) + 
\nonumber \\
&&\ \ \ \ \ \frac{2\varepsilon \Pi }{S_{2}}\frac{\partial }{\partial \xi }%
\left( \xi P\right) +\left[ F(0)-F(r)\right] \frac{\partial ^{2}P}{\partial
\xi ^{2}}.  \label{dpmod}
\end{eqnarray}
This equation is not strictly closed because the evaluation of $%
S_{2}(r)=\int \xi ^{2}P(\xi ,r)d\xi $ requires the knowledge of $P$ in the
whole range of $\xi $, not only in the large $\xi $ asymptotic where (\ref
{dpmod}) holds. However, $S_{2}$ is a much simpler object to deal with,
either theoretically or experimentally, than conditional averages. Besides,
much and important information can be gathered without knowledge of its
explicit expression as shown now. Let us consider now the stationary
solutions of (\ref{dpmod}), in which case the rate of energy injection $%
\varepsilon ^{\prime }$ equals the rate of energy dissipation $\varepsilon $%
. The forcing correlation is then written as 
\begin{equation}
F(r)=2\varepsilon \chi (r/L_{c}),  \label{fnondim}
\end{equation}
where $\chi $ is an even, dimensionless function that rapidly decays for
arguments larger than one, also satisfying $\chi (0)=1$. For $r$ large
compared to the forcing correlation length $L_{c}$ one then has $F(r)\simeq
0 $, and $S_{2}\simeq 2\left\langle u^{2}\right\rangle $, which is $r$
independent. With these considerations, (\ref{dpmod}) immediately leads to
the Gaussian solution $P(\xi )=\exp (-\xi ^{2}/(2S_{2}))$, as it must since
the two velocities involved are statistically independent for the $r$'s
considered. Much more interesting is the situation $r<<L_{c}$, in which case
one can write $F(r)=2\varepsilon \left[ 1+\chi ^{\prime \prime
}(0)\,(r/L_{c})^{2}/2+O((r/L_{c})^{4})\right] $, with $\chi ^{\prime \prime
}(0)<0$. If $\xi ^{2}P\rightarrow 0$ at large $\xi $, the immediate integral
of (\ref{dpmod}) gives 
\begin{equation}
\ln P=\frac{L_{c}^{2}}{r^{2}\chi ^{\prime \prime }(0)}\left[ \frac{\xi ^{3}}{%
6\varepsilon }\frac{\partial \ln S_{2}}{\partial r}+\frac{\xi ^{2}\Pi }{S_{2}%
}\right] .  \label{pos}
\end{equation}
If in the range of $r$ considered $S_{2}(r)$ behaves as a power law, (in
fact, $S_{2}(r)\sim r^{2}$ for $r$ smaller than the dissipation scale, as
seen above, and $S_{2}(r)\sim r$ in the inertial range \cite{got}) we have $%
\partial \ln S_{2}/\partial r\sim r^{-1}$. Taking into account that $\chi
^{\prime \prime }(0)<0$, it then results that the leading behavior for large 
$\xi $ is $P\sim \exp [-L_{c}^{2}\xi ^{3}/(\varepsilon r^{3})]$. This
solution is of course valid only for $\xi >0$. For $\xi <0$ then, $\xi ^{2}P$
cannot approach zero as $\xi \rightarrow -\infty $, and so $P$ must behave
as 
\begin{equation}
P=\frac{C(r)}{\xi ^{2}},  \label{neg}
\end{equation}
where $C(r)$ is an unknown, positive function of $r$.

\subsubsection{Pdf of velocity derivative}

Let us consider now the pdf of velocity derivative at the origin, $G\left[
u\right] =\left. \partial u/\partial x\right| _{x=0}$, so that $h(\eta
)=\delta G/\delta u(\eta )=-\delta ^{\prime }(\eta )$, where the prime
indicates derivative with respect to the argument. From Eqs. (\ref{pdfgen})
and (\ref{forcing}) we readily obtain 
\begin{eqnarray*}
\frac{\partial P}{\partial t} &=&\frac{\partial }{\partial \xi }\left[
\left( \xi ^{2}+\left\langle u\frac{\partial ^{2}u}{\partial x^{2}}|\xi
\right\rangle -\nu \left\langle \frac{\partial ^{3}u}{\partial x^{3}}|\xi
\right\rangle \right) P\right] \\
&&-\frac{1}{2}F^{\prime \prime }\left( 0\right) \frac{\partial ^{2}P}{%
\partial \xi ^{2}}.
\end{eqnarray*}

We now write, using expression (\ref{finalmodel}) with $m=2$, 
\begin{eqnarray*}
\left\langle u\frac{\partial ^{2}u}{\partial x^{2}}|\xi \right\rangle
&=&\left. \frac{\partial ^{2}}{\partial x^{2}}\left\langle u\left( 0\right)
u\left( x\right) |\xi \right\rangle \right| _{x=0} \\
&\rightarrow &\xi ^{2}\frac{\left\langle u\frac{\partial u}{\partial x}%
\right\rangle \left\langle \frac{\partial u}{\partial x}\frac{\partial ^{2}u%
}{\partial x^{2}}\right\rangle }{\left\langle \left( \frac{\partial u}{%
\partial x}\right) ^{2}\right\rangle ^{2}}=0,
\end{eqnarray*}
where the zero comes from the assumed statistical homogeneity. Using now
expression (\ref{finalmodel}) with $m=1$ we have, using also the condition
of homogeneity, 
\begin{eqnarray*}
\left\langle \frac{\partial ^{3}u}{\partial x^{3}}|\xi \right\rangle
&=&\left. \frac{\partial ^{3}}{\partial x^{3}}\left\langle u\left( x\right)
|\xi \right\rangle \right| _{x=0} \\
&\rightarrow &-\xi \frac{\left\langle \left( \frac{\partial ^{2}u}{\partial
x^{2}}\right) ^{2}\right\rangle }{\left\langle \left( \frac{\partial u}{%
\partial x}\right) ^{2}\right\rangle }.
\end{eqnarray*}
so that, calling $\kappa \equiv \nu \left\langle \left( \partial
^{2}u/\partial x^{2}\right) ^{2}\right\rangle \left\langle \left( \partial
u/\partial x\right) ^{2}\right\rangle ^{-1}=\varepsilon ^{-1}\left\langle
\left( \nu \partial ^{2}u/\partial x^{2}\right) ^{2}\right\rangle $, we have
for the statistically stationary case, using also the expression (\ref
{fnondim}) for the forcing, 
\[
\frac{\partial }{\partial \xi }\left[ \left( \xi ^{2}+\kappa \xi \right)
P-\varepsilon \frac{\chi ^{\prime \prime }\left( 0\right) }{L_{c}^{2}}\frac{%
\partial P}{\partial \xi }\right] =0, 
\]
and analogously to the velocity difference case we obtain for $\xi >0$%
\[
\ln P=\frac{L_{c}^{2}}{\chi ^{\prime \prime }\left( 0\right) }\left( \frac{%
\xi ^{3}}{3\varepsilon }+\frac{\kappa \xi ^{2}}{2\varepsilon }\right) , 
\]
and, for $\xi <0$, 
\[
P\sim \xi ^{-2}. 
\]

All these results for the Burgers equation reproduce well the behavior
obtained theoretically and numerically in the references cited above.

\section{Conclusions}

We have presented a simple closure for the asymptotic pdf evolution equation
of rather generic turbulence problems. We have presented results for the
simplest possible problems in turbulence, for which a rather detailed
knowledge exists. For the paradigmatic problem of Navier-Stokes turbulence
the application of the formalism is more involved, and has been applied in a
particular version in\cite{fom}, where very reasonable results were
obtained, although comparison with detailed pdf 's is more difficult.

{\large Acknowledgements}

This research was supported by grants of the Consejo Nacional de
Investigaciones Cient\'{i}ficas y T\'{e}cnicas (PEI 6004) and the University
of Buenos Aires (PID X-106).

\end{document}